\documentstyle[aps,prd,epsf]{revtex}
\begin{document}
\newcommand{\ppp}{\partial}
\newcommand{\V}{\nabla}
\newcommand{\EEE}{\epsilon}
\newcommand{\nnb}{\nonumber}
\newcommand{\s}[1]{{\scriptscriptstyle{#1}}}
\newcommand{\ssr}[1]{\scriptscriptstyle{\rm \, #1}}
\newcommand{\real}{\mbox{${\rm I\!R}$}}
\newcommand{\bm}[1]{ \mbox{\boldmath{$#1$}}  }
\newcommand{\tfrac}[2]{ \textstyle\frac{#1}{#2} }
\newcommand{\kp}{\kappa}
\newcommand{\tA}{\tilde{a}}
\newcommand{\bU}{\bar{\cal U}}

%
%
\leftline{\epsfbox{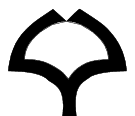}}
\vspace{-10.0mm}
\thispagestyle{empty}
{\baselineskip-4pt
 \font\yitp=cmmib10 scaled\magstep2
 \font\elevenmib=cmmib10 scaled\magstep1 \skewchar\elevenmib='177
 \leftline{\baselineskip20pt
           \hspace{12mm} 
           \vbox to0pt
             { {\yitp\hbox{Osaka \hspace{1.5mm} University} }
               {\large\sl\hbox{{Theoretical Astrophysics}} }
              \vss}
          }
}
%
%
{\baselineskip0pt
 \rightline{\large\baselineskip14pt\rm\vbox
            to20pt{\hbox{OU-TAP-111}
                   \hbox{UTAP-358}
              \vss}
           }
}
\vskip15mm
%
%
\begin{center}
{\large\bf False Vacuum Decay with Gravity in Non-Thin-Wall Limit}
\end{center}
\begin{center}
{\large
   Uchida Gen and  Misao Sasaki\\
\bigskip
\sl{ Department of Earth and Space Science, Graduate School of Science,\\
          Osaka University, Toyonaka 560-0043, Japan\\
              \vskip2.5mm    and\\  \vskip2.5mm
     Department of Physics, The University of Tokyo, Tokyo 113-0033, Japan
   }
}
\end{center}
\begin{abstract}
We consider a wave-function approach to the false vacuum decay
with gravity and present a new method to calculate the tunneling
amplitude under the WKB approximation. The result agrees with
the one obtained by the Euclidean path-integral method, but gives a
much clearer interpretation of an instanton (Euclidean bounce
solution) that dominates the path integral. In particular, our method is
fully capable of dealing with the case of a thick wall
with the radius of the bubble comparable to the radius of the instanton,
thus surpassing the path-integral method whose use
can be justified only in the thin-wall and small bubble radius limit.
The calculation is done by matching two WKB wave functions,
one with the final state and another with the initial state,
with the wave function in the region where the scale factor of the metric is
sufficiently small compared with the inverse of the typical
energy scale of the field potential at the tunneling.
The relation of the boundary condition on our wave function for the
false vacuum decay with Hartle-Hawking's no-boundary boundary condition
and Vilenkin's tunneling boundary condition on the wave function of the
universe is also discussed.
\end{abstract}
\vskip1cm
\section{Introduction}
In the previous literature, the false vacuum decay with gravity
was investigated, in essence, by a naive extrapolation of
well-established methods used for the Minkowski background\cite{CD}.
There are two methods to calculate the decay rate of false vacuum,
one using the Euclidean path integral and the other constructing
a WKB wave function. However, in both approaches such
extrapolation is valid only for a tunneling potential
$V(\phi)$ satisfying the thin-wall condition (we shall come back
to this point below).

The false vacuum decay in quantum field theory on the Minkowski
background is analyzed in the semiclassical way (i.e., in the
WKB approximation) as follows\cite{C}.
First, the Euclidean classical equation of motion for the field $\phi$
(we consider a real scalar field throughout the paper) is
solved by assuming the $O(4)$ symmetry on the field configuration with
the boundary condition that the solution is regular at the origin
and asymptotically approaches the false vacuum at infinite radius.
There will be two such solutions; one trivially sitting at
the false vacuum and the other that leaves the origin from a field value
near the true vacuum and goes over the potential barrier.
We call the former the false vacuum instanton and the latter
the bounce instanton. We can take either the
path-integral approach or the wave-function approach.

In the path-integral approach,
we sum up the contribution of instantons and fluctuations around them to
obtain the tunneling amplitude or the decay rate of false vacuum.
If we denote the Euclidean action of the false vacuum instanton
by $S_F$ and that of the bounce instanton by $S_B$, the decay rate
is given by $\Gamma\sim e^{-(S_B-S_F)}$. In passing, we note that
in the Minkowski background it is customary to set $V(\phi_F)=0$ so
that $S_F=0$. But one can take a different choice, $V(\phi_F)\neq0$,
in which case both $S_F$ and $S_B$ would diverge but the difference
is well-defined and independent of the value of $V(\phi_F)$.

In the wave-function approach, we construct a one-parameter family of
3-dimensional field configurations from the everywhere-false-vacuum
configuration to the configuration with the true vacuum bubble
about to expand in the false vacuum sea (the critical bubble
configuration). Then, exponentiating the action of these configurations
we obtain the WKB wave function(al) describing the tunneling
of the field from the false vacuum state to the state of the
critical bubble \cite{mdt}.
Then the tunneling amplitude is obtained by taking the square of the
ratio of the final state wave function to the initial state wave
function, which is also given by $e^{-(S_B-S_F)}$.

Since these two different approaches give the same result,
and since there is one-to-one correspondence between a classical
solution and a WKB wave function, we may justify the interpretation
of the bounce instanton as describing the tunneling process and its
analytic continuation to the Lorentzian space-time as describing the
classical configuration of the field after false vacuum decay.

\begin{figure}
\centerline{\epsfxsize=6cm\epsfbox{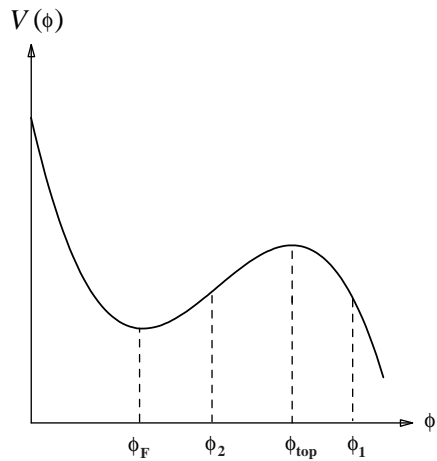}}
\caption{ A schematic picture of the potential $V(\phi)$
of a scalar field discussed in the text.
}
\label{fig:pt}
\end{figure}

A trouble arises when one attempts to extrapolate the above procedure
to the case of false vacuum decay with gravity.
First, the scalar field equation is now coupled with the
Friedmann equation that governs the scale factor of
the $O(4)$ symmetric metric. Second, and most problematically,
the 4-dimensional Euclidean space becomes inevitably compact with
topology of $S^4$ when gravity is taken into account
(we assume $V(\phi_F)>0$ where $\phi_F$ is the field
value at the false vacuum and a potential of the shape depicted in
Fig.~\ref{fig:pt}), i.e., not $\real^4$ as in the case of flat
space-time.
Of course a bounce solution still exists in the presence of gravity,
first obtained by Coleman and DeLuccia\cite{CD}.
However, there will be no bounce solution that admits
a 3-dimensional slice on which the field is {\em asymptotically} at
false vacuum everywhere; an instanton leaving, say, the north pole of
$S^4$ from the true vacuum side does not reach the false vacuum
when arriving even at the south pole.
Furthermore, the standard dilute gas approximation
to obtain the decay rate in the path-integral approach,
in which the 4-volume occupied by a vacuum bubble is assumed to be
negligible compared with the whole Euclidean 4-volume, fails.

Nevertheless, there is one special case in which the extrapolation
can be almost justified. It is the thin-wall limit\cite{JS1}.
In this case, the field $\phi$, which is sitting on the true vacuum side
of the barrier at the north pole,
varies abruptly at the position of the bubble wall sharply
located in the northern hemisphere and $\phi=\phi_F$ everywhere outside
the bubble. Then, the solution admits
a maximal 3-surface on which the extrinsic curvature of the 3-surface
vanishes and $\phi=\phi_F$ everywhere. Through this surface, we can
analytically continue to the Lorentzian solution that describes the
false vacuum state. The critical bubble configuration that describes
the moment of bubble nucleation is also a maximal 3-surface, and
analytic continuation through the surface gives the space-time with
an expanding bubble.  Thus we obtain a one-parameter family of
3-dimensional configurations that interpolates between the false vacuum
and critical bubble configurations.

However, only a very restricted class of potentials
admits bounce instantons in the thin-wall limit. Namely, the ratio of
the mass scale $m$ of the curvature at the top of
the potential barrier, $m^2=-d^2V/d\phi^2(\phi_{top})$,
to the typical mass scale $M$ of the potential energy density
must be very large (rigorously speaking, it must be infinitely
large in the exact thin-wall limit).
In other words, the barrier must be extremely sharply peaked.
We note that this is somewhat different from the definition of the
thin-wall limit originally discussed by Coleman and DeLuccia\cite{CD}.
In our definition, since the Euclidean 4-volume is proportional
to $(GM^{4})^{-2}$ while the 4-volume of a bubble is
 $O(m^{-4})$, the thin-wall limit implies small bubble radius,
provided $M$ is smaller than the Planck scale, i.e., $GM^2<1$.
Field potentials with such a feature are apparently not general.
However, previous discussions on the false vacuum decay
with gravity have been relied on the picture of a bounce instanton
having the feature of the thin-wall limit.
For example, it is widely believed that the formula $\Gamma\sim
e^{-(S_B-S_F)}$ is still valid even when the bounce solution
does not have the thin-wall feature at all.
Furthermore, the interpretation of such a bounce as describing the
tunneling process and describing the classical evolution from the
critical bubble state has been adopted without serious
considerations.

In this paper, we propose a new method to calculate the wave function
describing false vacuum decay with gravity, which is not restricted
to the thin-wall limit. Our method gives a picture of tunneling
substantially different from that in the Minkowski background,
but nevertheless our wave function interpolates between the false vacuum
state, which is described by the trivial instanton sitting at false
vacuum, and the state at the bubble nucleation, which is a maximal
3-slice containing the critical bubble of the bounce solution.

The paper is organized as follows. In Section II, we first review
the mini-superspace Wheeler-DeWitt equation for $O(4)$ symmetric
configurations and the WKB approximation. Then we describe our
method. We construct a one-parameter family of spatial configurations
that interpolates between the false vacuum state and the critical
bubble state. As we are interested in the WKB approximation,
we consider a classical path connecting these two states.
As pointed out in the above, no `single' classical solution
admits such a path. However, a unique feature of false
vacuum decay with gravity is that one can match the `two'
instantons (the false vacuum instanton and the bounce instanton)
smoothly across the `south pole' of each instanton.
In section III, by solving the Wheeler-DeWitt equation we explicitly
perform this matching and construct the wave function that
contains both the false vacuum and critical bubble states, hence
describes the false vacuum decay with gravity for a wide class of
tunneling potentials.
It is then straightforward to calculate the tunneling amplitude.
The result recovers the Minkowski result if we take
the zero-gravity limit $G\to0$, and also agrees with the decay rate
obtained in the path-integral method with a naive extrapolation
of the formula, $\Gamma\sim e^{-(S_B-S_F)}$.
Our result supports the standard interpretation that
the bounce solution, despite the fact it does not contain
the false vacuum configuration at all, does describe
the classical evolution after false vacuum
decay by analytic continuation through the critical bubble
configuration.
Section IV is devoted to discussions. A covariant formulation
of the WKB wave function for tunneling is recapitulated in
Appendix A.

In the rest of the paper, we keep $\hbar$ and $G$ explicit
to clarify the WKB order as well as the effect of
gravity.

\section{Formulation}
In this section, we describe our method to
construct a relevant classical path in the Euclidean regime
that determines the WKB tunneling wave function.

\subsection{Instantons and WKB approximation}
We consider the action,
\begin{equation}
  S=\int d^4x \sqrt{g}
        \left[ \frac{{\cal R}}{24\pi^2\kp}
               - \frac12 g^{\mu\nu} \ppp_\mu \phi \ppp_\nu \phi
               -V(\phi)
        \right],
\end{equation}
where $\kp:=2G/3\pi$, and the potential $V(\phi)$ as illustrated
in Fig.~\ref{fig:pt}.
We confine our attention to WKB wave functions that are described
by classical solutions having $O(4)$ symmetry.
Hence we consider the metric in the form,
\begin{equation}
  ds^2=-N^2(t)dt^2+a^2(t)d\Omega^2_{(3)}
\end{equation}
in which $d\Omega^2_{(3)}$ is the metric on a unit 3-sphere,
and $\phi=\phi(t)$.
In this case the action in the first-order form becomes
\begin{eqnarray}
S=\int dt L\,;\quad
L=P_a \dot a+P_\phi\dot\phi
-{\kappa N\over2a}\left(-P_a^2+{1\over(2\pi^2\kappa)a^2}P_\phi^2
-{a^2\over\kappa^2}Q(a,\phi)\right).
\end{eqnarray}
Then the Wheeler-DeWitt equation is written down as
\begin{equation}
  {\kp\over{2a}}
   \left[ \hbar^2\frac{\ppp^2}{\ppp a^2}
       +\hbar^2{p\over a}{\partial\over\partial a}
        - \frac{1}{(2\pi^2\kp)\,a^2}\hbar^2\frac{\ppp^2}{\ppp\phi^2}
        -\frac{a^2}{\kp^2} Q(a,\phi)
   \right] \Psi = 0,
\label{eq:WD}
\end{equation}
where
\begin{equation}
  Q(a,\phi)=1- (4\pi^2\kp)\, a^2 V(\phi),
\end{equation}
and $p$ is an arbitrary constant that determines
the operator-ordering; $p=1$ corresponds to making the ordering
covariant with respect to the superspace metric (see Appendix A).

It should be noted, however, that it is not always necessary to
take the $t=$const. 3-geometry as the argument of the wave function.
In fact, the critical bubble configuration that describes the moment
of bubble nucleation does not respect the $O(4)$ symmetry.
Nevertheless, we shall see below that the tunneling amplitude can be
calculated solely with the knowledge of the wave function for
$O(4)$ symmetric 3-geometries.

The construction of a WKB wave function describing
multi-dimensional quantum tunneling was discussed much
in the literature \cite{mdt} and reformulated in a covariant
manner in \cite{TSY}.
Given a one-parameter family of configurations that
satisfies the Euclidean equations of motion, the WKB wave function
along the family can be obtained with this method,
as recapitulated in Appendix A.
Following this method, we first express the wave function
in the form,
\begin{eqnarray}
\Psi=\exp[-W_0(a,\phi)/\hbar-W_1(a,\phi)+\cdots].
\end{eqnarray}
Then at the lowest WKB order, we have
\begin{equation}
  \left\{ \frac{\kp}{a}
           \left(\frac{\ppp W_0}{\ppp a}\right)^2
                   -\frac{1}{2\pi^2 a^3}
                          \left(\frac{\ppp W_0}{\ppp\phi}\right)^2
           \right\}
  -\frac{a}{\kp}Q(a,\phi) =0,
\label{eq:WKB}
\end{equation}
which is, of course, free from the operator-ordering ambiguity.

Introducing a parameter $\tau$ such that
\begin{equation}
   \frac{da(\tau)}{d\tau}:=
         -\frac{\kp}{a}  \frac{\ppp W_0}{\ppp a}
            \hskip5ex \mbox{and} \hskip5ex
   \frac{d\phi(\tau)}{d\tau}=:
        \frac{1}{2\pi^2 a^3} \frac{\ppp W_0}{\ppp\phi},
\label{eq:rel}
\end{equation}
we find
\begin{eqnarray}
{d\over d\tau}\,W_0\left(a,\phi\right)
=-{a\over\kappa}\,Q\left(a,\phi\right).
\label{eq:W0eq}
\end{eqnarray}
The Euclidean equations of motion
for $O(4)$ symmetric configurations, $(a(\tau),\phi(\tau))$,
with the choice of the lapse $N=1$
are obtained from Eqs.~(\ref{eq:WKB}), (\ref{eq:rel})
and (\ref{eq:W0eq}) as
\begin{eqnarray}
&& \dot a^2-(2\pi^2\kappa)a^2\dot\phi^2-Q(a,\phi)=0\,,
\nonumber\\
&&  \ddot\phi+3\frac{\dot{a}}{a}\dot{\phi} = \frac{dV}{d\phi}\,,
\nonumber\\
&&   a\ddot{a}+2\dot{a}^2+1=3Q(a,\phi).
\label{eq:eom}
\end{eqnarray}
The first equation is the (Euclidean) Friedmann equation
that corresponds to the energy integral of the remaining
two equations. The regularity of the solution
requires $\dot\phi=0$ at $a=0$.

Equation~(\ref{eq:W0eq}) readily gives
the relative amplitudes of the leading-order WKB wave function at
two different configurations at $\tau=\tau_1$ and $\tau_2$,
\begin{eqnarray}
\Psi(\tau_2)=\exp\left[-W_0(\tau_2;\tau_1)\right]\Psi(\tau_1)\,;
\quad
W_0(\tau_2;\tau_1)=-{1\over\kappa}\int^{\tau_2}_{\tau_1}
 d\tau'\,a(\tau')Q\left(a(\tau'),\phi(\tau')\right).
\label{eq:W0}
\end{eqnarray}
As is well-known, $W_0(\tau_2;\tau_1)$ is just the Euclidean
action integral of the system from the configuration at
$\tau_1$ to that at $\tau_2$.

\begin{figure}
\centerline{\epsfysize=5cm\epsfbox{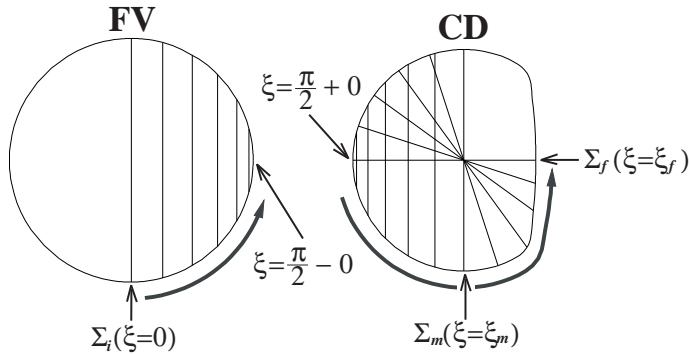}}
\caption{Schematic pictures of the false vacuum (FV) instanton
and the Coleman-DeLuccia (CD) instanton
with 2-dimensions suppressed.
The arrow indicates the direction of the increase in
the parameter $\xi$ defined in the text.}
\label{fig:fvcd}
\end{figure}

The solutions relevant to our discussion are the trivial solution
sitting at the false vacuum and the bounce instanton.
The trivial instanton is explicitly given by
\begin{eqnarray}
\left(a^{FV}(\tau),\phi^{FV}(\tau)\right)
=\left(H_{F}^{-1}\sin(H_{F}\tau), \phi_F \right)\,;
\quad \tau\in[0,\pi/H_F],
\label{eq:FVinst}
\end{eqnarray}
where $H_F=2\pi\sqrt{\kp V(\phi_F)}$.
It represents a Euclidean 4-sphere of radius $H_{F}^{-1}$
and we call it the false vacuum (FV) instanton hereafter.
The bounce instanton is known as the Coleman-DeLuccia (CD)
instanton \cite{CD,JS1}. We denote it by
$\left(a^{CD}(\tau),\phi^{CD}(\tau)\right)$ with
$\tau\in[\tau_1,\tau_2]$.
It also has the topology of $S^4$ but the field $\phi$
varies monotonically over the potential barrier as the
scale factor $a$ varies from zero to a maximum, then to zero.
For later convenience, we set
$\left(a^{CD}(\tau_1),\phi^{CD}(\tau_1)\right)=
\left(0,\phi_1\right)$
where $\phi_1$ is on the true vacuum side of the
barrier, and
$\left(a^{CD}(\tau_2),\phi^{CD}(\tau_2)\right)
=\left(0,\phi_2\right)$
where $\phi_2$ is on the false vacuum side of the barrier.
See Fig.~\ref{fig:pt}.
The FV and CD instantons embedded in $\real^5$
are schematically shown in
Fig.~\ref{fig:fvcd} with 2-dimensions suppressed.

\subsection{One-parameter family of tunneling configurations}

As we noted in Introduction, the CD instanton itself does not
admit a maximal 3-surface on which $\phi=\phi_F$ everywhere,
unless the potential satisfies the thin-wall
condition. Hence the CD instanton alone cannot describe the
tunneling wave function. It is then natural to suppose that
the FV instanton plays a role as well. In fact, there is already
a hint in the formula for the decay rate,
$\Gamma\sim e^{-(S_B-S_F)/\hbar}$, valid in the thin-wall limit,
where the action of the FV instanton comes into play.

We therefore consider a possibility to construct a
one-parameter family of spatial configurations that
connect the false vacuum state and the critical bubble
state by matching these two instantons somehow.
To look for this possibility, we have to keep
in mind that the WKB approximation
requires us that such a path should satisfy the
classical equations of motion {\em almost everywhere}.
But this is the crucial point;
it is so not everywhere but almost everywhere.
In any calculation of a wave function under the
WKB approximation, there can be configurations
of measure zero that violate the WKB condition,
such as turning points of a classical solution,
but their existence does not invalidate the calculation
if one appropriately performs the matching,
e.g., by the method of asymptotic matching.
Then we realize that the two instantons
may be actually matched through the point $a=0$
where the 3-geometry ceases to exist, hence
can be regarded as a turning point of the
classical solutions.

With the above considerations in mind,
we combine and re-organize the $O(4)$ symmetric families of
configurations of the two instantons discussed in the previous
subsection to a one-parameter family of configurations
which adequately describes the false vacuum decay.
We denote this one-parameter family by
$(\bar{h}_{ij}(x^i;\xi),\bar{\phi}(x^i;\xi))$
where $\xi$ is a non-dimensional parameter.
The parameter $\xi$ is assumed to run
through a range $[0,\xi_f]$ with the initial ($\xi=0$)
and final ($\xi=\xi_f$) configurations given by the
the maximal 3-slice of the FV instanton and
the critical bubble configuration of the CD instanton,
respectively.

The initial configuration, denoted by $\Sigma_i$ in
 Fig.~\ref{fig:fvcd}, is given by
$H_F\tau=\pi/2$ of Eq.~(\ref{eq:FVinst}).
We thus set $\xi=H_F\tau-\pi/2$ until
the scale factor $a$ vanishes,
\begin{equation}
  \left(\bar{h}_{ij}(x^i;\xi),\bar{\phi}(x^i;\xi)\right)=
\left(H_{F}^{-2}\sin^2(\xi+\frac{\pi}{2})d\Omega_{(3)}^2,\phi_F\right);
\quad \xi\in[0,{\pi/  2}).
\end{equation}

We match $a=0$ of the above to the point
$\left(a^{CD},\phi^{CD}\right)=(0,\phi_2)$ of the CD instanton,
where $\phi_2$ is the field value at the `south pole',
see Figs.~\ref{fig:pt} and \ref{fig:fvcd}.
Note that the values of $\phi$ are different at these two
points where the WKB breaks down. However, as we shall see in
the next section,
this will not be an obstacle since the wave function
will be independent of $\phi$ in the vicinity of $a=0$.

For $\xi>\pi/2$, we
take the $O(4)$ symmetric configurations of the CD instanton
up to the slice $\Sigma_m$ shown in Fig.~\ref{fig:fvcd},
where $\Sigma_m$ is the $O(4)$ symmetric slice
at $\dot a^{CD}=0$ where the 3-volume is maximum,
which we denote by
$(\bar{h}_{ij},\bar{\phi})=(H_{m}^{-2}d\Omega^2_{(3)},\phi_m)$.
Thus
\begin{eqnarray}
\left(\bar{h}_{ij}(x^i;\xi),\bar{\phi}(x^i;\xi)\right)=
\left(a^{CD}(\xi)d\Omega_{(3)}^2,\phi^{CD}(\xi)\right);
\quad \xi\in({\pi\over 2},\xi_m],
\end{eqnarray}
where
\begin{eqnarray}
\left(a^{CD}({\pi/2}),\phi^{CD}({\pi/2})\right)
=(0,\phi_2),\quad
\left(a^{CD}(\xi_m),\phi^{CD}(\xi_m)\right)
=(H_m^{-1},\phi_m).
\end{eqnarray}

For $\xi>\xi_m$, we slice the CD instanton in such a way
that all of the configurations contain the common 2-sphere
the slice $\Sigma_m$ and the final slice $\Sigma_f$ of the critical
bubble configuration intersect.
These slices do not respect the $O(4)$ symmetry but
are analogous to the static slices
of the de Sitter space of radius $H^{-1}$:
\begin{equation}
   h_{ij}(x^i)=\frac{dr^2}{1-H^2 r^2}+r^2d\Omega^2_{(2)}\,.
\end{equation}
As mentioned above, we label the final configuration
by $\xi=\xi_f$. Thus
\begin{eqnarray}
\left(\bar{h}_{ij}(\xi_f),\bar{\phi}(\xi_f)\right)
=(\hbox{the critical bubble 3-geometry}).
\end{eqnarray}

As it is clear from the above construction,
the lapse function $N$ vanishes at $\Sigma_m$, and the geometry
is degenerate there.
However, this will not be a problem since we have started from the
Wheeler-DeWitt equation (\ref{eq:WD}), in which the dynamical
variables are $(h_{ij}(x^i),\phi(x^i))$ of a 3-geometry.
Hence, although the lapse function plays a role when
a Euclidean classical solution is considered,
the vanishing of it is irrelevant to our
discussion (in this connection, see \cite{FMP}).
Furthermore, as the relative magnitude of the WKB wave function
at two different configuration is simply determined by
the action integral between them, and it is independent of
deformation of a path between them (as long as
the path lies on the space of the classical solution that
dominates the wave function),
the contribution from the upper-left quadrant of the CD instanton
in Fig.~\ref{fig:fvcd} swept both by the configurations
between $\xi\in(\pi/2,\xi_m]$ and $\xi\in[\xi_m,\xi_f]$
cancel each other and only the action of the
lower hemisphere determines the relative magnitude
of the wave function at $\xi=\pi/2$ and $\xi=\xi_f$.
Thus, although the values of the wave function
for configurations in the range $\xi\in(\xi_m,\xi_f)$
are difficult to obtain, they are not needed in the
calculation of the tunneling amplitude.

\section{Tunneling wave function}

Let us now construct the tunneling wave function.
Since we are interested in the false vacuum decay,
we want the wave function to describe the {\em expanding} universe
after tunneling.
Thus the appropriate boundary condition for
the wave function is the tunneling boundary
condition, which
demands the wave function describing the classical universe
after tunneling to have a positive eigenvalue for
the momentum operator $P_a=-i\hbar\partial/\partial a$ \cite{V}.

In accordance with the construction of the classical path given
in the previous section, we assume the universe after tunneling
is described by analytic continuation of the CD instanton
through the critical bubble configuration.
We choose the moment of nucleation ($\xi=\xi_f$) as $t=\tau=0$.
Thus we set
\begin{equation}
  \Psi(\xi>\xi_f)= N
    \exp \left[{i\over\hbar}
 \int^{t(\xi)}_{0}dt'L^{CD}(t')
                     +i \frac{\pi}{4}
                   \right],
\end{equation}
where $N$ is a normalization constant and $L^{CD}(t)$ is the
Lagrangian of the Lorentzian CD solution. The phase $\pi/4$
is inserted for convenience, and
we have neglected the prefactor arising from the next WKB order.
Then, the standard WKB connection formula gives
the under-barrier wave function
for $\xi\in(\pi/2,\xi_f]$ as
\begin{equation}
    \Psi(\pi/2<\xi<\xi_f)
  =N\exp\left[{1\over\hbar}
\int^{0}_{\tau(\xi)}d\tau'L_E^{CD}(\tau') \right]
  +\frac{i}{2}
   N\exp\left[-{1\over\hbar}
\int^{0}_{\tau(\xi)}d\tau'L_E^{CD}(\tau')\right],
\label{eq:WV+}
\end{equation}
where $L_E^{CD}(\tau)$ is the Euclidean Lagrangian of the CD solution.

For $\xi\in(\pi/2,\xi_m]$ where the 3-geometries are $O(4)$ symmetric,
we have from Eq.~(\ref{eq:W0}),
\begin{eqnarray}
\int^{0}_{\tau(\xi)}d\tau'L_E^{CD}(\tau')
&&=W_0(\xi_f;\pi/2)-W_0(\xi;\pi/2)
\nonumber\\
&&={S_B\over2}-W_0(\xi;\pi/2),
\label{Psicd}
\end{eqnarray}
where $S_B$ is the total action of the CD bounce instanton
and
\begin{eqnarray}
W_0(\xi;\pi/2)&&=-{1\over\kappa}\int_{\tau(\pi/2)}^{\tau(\xi)}d\tau'\,
a^{CD}(\tau')Q\left(a^{CD}(\tau'),\phi^{CD}(\tau')\right)
\nonumber\\
&&=:W_0^{CD}(\xi),
\end{eqnarray}
where $a^{CD}(\tau(\pi/2))=0$. Hence
\begin{eqnarray}
\Psi(\pi/2<\xi<\xi_m)
=A\exp\left[-{1\over\hbar}W_0^{CD}(\xi)\right]
+B\exp\left[{1\over\hbar}W_0^{CD}(\xi)\right].
\label{WKB>}
\end{eqnarray}
where the coefficients $A$ and $B$ are given by
\begin{eqnarray}
A=Ne^{S_B/2\hbar}\,,\quad B={i\over2}Ne^{-S_B/2\hbar}\,.
\label{ABvalue}
\end{eqnarray}
Likewise, the WKB wave function for $\xi\in(0,\pi/2)$ is expressed as
\begin{eqnarray}
\Psi(0<\xi<\pi/2)
=C\exp\left[-{1\over\hbar}W_0^{FV}(\xi)\right]
+D\exp\left[{1\over\hbar}W_0^{FV}(\xi)\right],
\label{WKB<}
\end{eqnarray}
where $C$ and $D$ are constants and
\begin{eqnarray}
W_0^{FV}(\xi):=&&W_0(\pi/2;\xi)
\nonumber\\
=&&-{1\over\kappa}\int_{\tau(\xi)}^{\pi/H_F}d\tau'
a^{FV}(\tau')Q\left(a^{FV}(\tau'),\phi_F\right),
\end{eqnarray}
where $a^{FV}(\tau)$ is given by Eq.(\ref{eq:FVinst}).

We consider the matching of the above wave functions in a
region $a\sim0$.
Noting that $V(\phi)\ge V(\phi_F)$
in the vicinity of $\xi=\pi/2$,
it is convenient to introduce a re-scaled (non-dimensional)
scale factor,
\begin{equation}
   \tA:= \epsilon \frac{a}{\ell_{pl}}\,,
\end{equation}
where the parameter $\epsilon$ is defined by
\begin{equation}
   \epsilon:=2\pi\sqrt\frac{\hbar^3 V(\phi_F)}{m_{pl}^4}\,,
\end{equation}
and $\ell_{pl}:=\sqrt{\hbar\kp}=\sqrt{2\hbar G/3\pi}$
is the Planck length and $m_{pl}:=\sqrt{\hbar/\kp}=\sqrt{3\pi\hbar/2G}$
the Planck mass. We assume the potential energy scale is
much below the Plank scale, $\epsilon \ll 1$.

In the region $a \ll 1/(2\pi\sqrt{\kp V(\phi_F)})$,
which corresponds to $\tA \ll 1$,
$Q(a,\phi)$ is independent of $\phi$; $Q\approx 1$, but
the WKB approximation is valid as long as $\tA\gg \epsilon$.
Hence, $W_0^{CD}(\xi)$ and $W_0^{FV}(\xi)$
in this region reduce to the same form,
\begin{eqnarray}
W_0^{CD}(\xi)=W_0^{FV}(\xi)
=-{a^2(\xi)\over2\kappa}
=-\hbar\frac{\tA^2(\xi)}{2\epsilon^2}\,.
\end{eqnarray}
The prefactor due to the first-order WKB correction
is calculated to be (see Appendix A)
\begin{equation}
     \exp({-W_1})=\tA^{-\frac{p+1}{2}}\times\mbox{const.}.
\end{equation}
Although we are not interested in this first order correction,
we recover it in the region $\epsilon\ll\tA\ll 1$
to show explicitly that the matching we perform
below is independent of the operator-ordering.
Thus the WKB wave function in the region $\epsilon\ll\tA\ll 1$
becomes
\begin{eqnarray}
   \Psi(\xi>\pi/2) &=&
       A\tA^{-\frac{p+1}{2}}
            \exp\left[+\frac{\tA^2(\xi)}{2\epsilon^2}\right]
     + B\tA^{-\frac{p+1}{2}}
            \exp\left[-\frac{\tA^2(\xi)}{2\epsilon^2}\right],
   \label{eq:WKBwf>}
\end{eqnarray}
and
\begin{eqnarray}
   \Psi(\xi<\pi/2) &=&
  C\tA^{-\frac{p+1}{2}}
            \exp\left[+\frac{\tA^2(\xi)}{2\epsilon^2}\right]
   + D\tA^{-\frac{p+1}{2}}
            \exp\left[-\frac{\tA^2(\xi)}{2\epsilon^2}\right].
  \label{eq:WKBwf<}
\end{eqnarray}

\begin{figure}[t]
\centerline{\epsfxsize=8cm\epsfbox{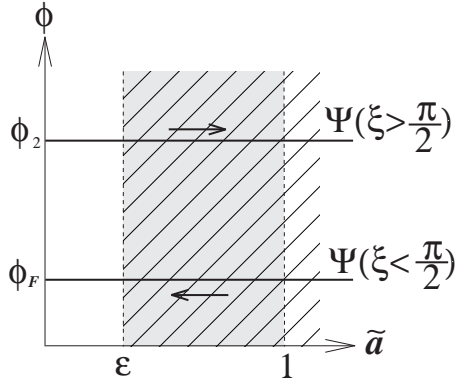}}
\caption{A picture of the superspace $(\tA,\phi)$
near the matching region.
The thick lines are the one-parameter family of configurations
$(\bar{h}_{ij}(\xi),\bar{\phi}(\xi))$ in the
range $\xi\in[0,\pi/2)$ and $\xi\in(\pi/2,\xi_f]$.
The arrow indicates the direction of increase in $\xi$ along the family.
The WKB approximation is valid in the hatched region
and the wave function (\protect{\ref{eq:WDwfapp}}) is valid
in the grayed region. The overlapping region
is where the matching is performed.}
\label{fig:phia}
\end{figure}

To match these two wave functions,
we directly solve the Wheeler-DeWitt equation
in the region $0 < \tA \ll 1$ and compare the coefficients
in the region $\epsilon\ll\tA\ll1$. See Fig.~\ref{fig:phia}.
In terms of $\tA$, the Wheeler-DeWitt equation (\ref{eq:WD})
reduces to
\begin{equation}
 \left[\frac{\ppp^2}{\ppp\tA^2}+\frac{p}{\tA}\frac{\ppp}{\ppp\tA}
           -\frac{\tA^2}{\epsilon^4}\right]\Psi=0.
\end{equation}
The general solution is
\begin{equation}
  \Psi(\tA,\phi) =
      \pi^{1/2} E \tA^{2\nu} I_{\nu}(\frac{\tA^2}{2\epsilon^2})
     +\pi^{-1/2}F \tA^{2\nu} K_{\nu}(\frac{\tA^2}{2\epsilon^2}),
               \label{eq:WDwf}
\end{equation}
where $\nu=(1-p)/4$, $E$ and $F$ are constants, and
$I_\nu(z)$ and $K_\nu(z)$ are the modified Bessel functions.
The asymptotic form of this wave function at $\tA\gg\epsilon$
is given by
\begin{equation}
  \Psi(\tA,\phi)
    = \left[
        \epsilon E \tA^{-\frac{p+1}{2}}
                \exp\left[+\frac{\tA^2}{2\epsilon^2}\right]
      + \epsilon(E e^{(\nu+\frac12)\pi i} + F) {\tA^{-\frac{p+1}{2}}}
                \exp\left[-\frac{\tA^2}{2\epsilon^2}\right]
      \right] (1+O(\frac{2\epsilon^2}{\tA^2})).
 \label{eq:WDwfapp}
\end{equation}

Since Eqs.~(\ref{eq:WKBwf>}) and (\ref{eq:WKBwf<}) have exactly the
same $\tA$ dependence, the comparison of them with
(\ref{eq:WDwfapp}) readily gives
\begin{eqnarray}
A=\epsilon E=C\,,\quad
B=\epsilon(E e^{(\nu+\frac12)\pi i}+F)=D\,,
\end{eqnarray}
independently of the choice of the operator-ordering.

\begin{figure}
\centerline{\epsfxsize=12cm\epsfbox{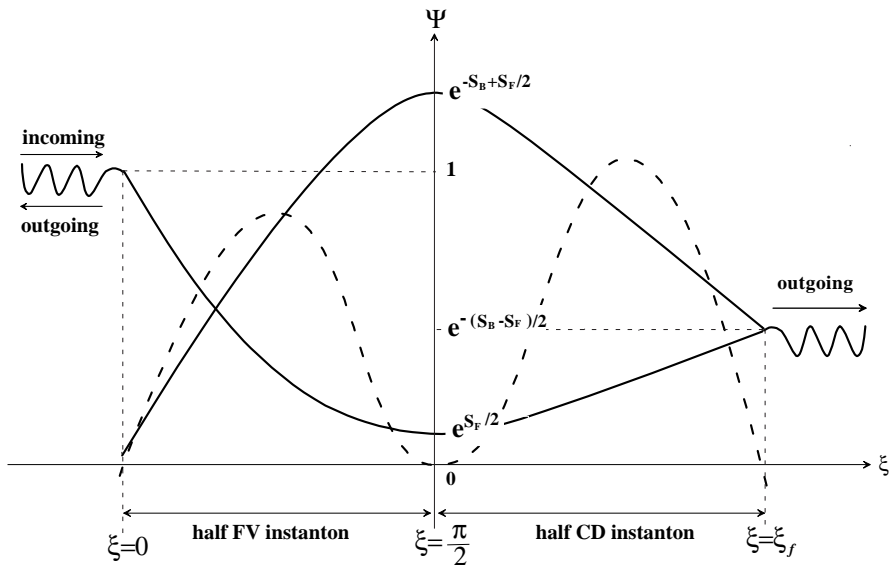}}
\caption{
A schematic behavior of the tunneling wave function.
The point $\xi=\pi/2$ is where the scale factor $a$
vanishes, and towards both right and left
directions from there the 3-volume increases.
The dotted curve indicates the effective potential barrier.}
\label{fig:wv2}
\end{figure}

Plugging this result into Eq.~(\ref{WKB<})
and noting Eq.~(\ref{ABvalue}), we find
the wave function for $\xi\in(0,\pi/2)$ as
\begin{equation}
  \Psi(0<\xi<\pi/2)=
        N e^{S_B/2\hbar} \exp\left[-{1\over\hbar}W_0^{FV}(\xi)\right]
  +\frac{i}{2}N e^{-S_B/2\hbar}
\exp\left[{1\over\hbar}W_0^{FV}(\xi)\right].
\end{equation}
Noting the fact $W_0^{FV}(0)=S_F/2$, we connect this to the Lorentzian
region $\xi<0$ of the false vacuum to obtain
\begin{eqnarray}
    \Psi(\xi<0)&=&
     N \exp\left[{1\over2\hbar}(S_B-S_F)\right]
   \left\{
        \exp\left[{i\over\hbar}\int^{t(\xi)}_{0} dt'L^{FV}(t')\right]
      + \exp\left[-{i\over\hbar}\int^{t(\xi)}_{0} dt'L^{FV}(t')\right]
   \right\}  \nonumber \\
  &&+\frac{i}{2} N\exp\left[-{1\over2\hbar}(S_B-S_F)\right]
   \left\{ 
        \exp\left[{i\over\hbar}\int^{t(\xi)}_{0} dt'L^{FV}(t')\right]
      + \exp\left[-{i\over\hbar}\int^{t(\xi)}_{0} dt'L^{FV}(t')\right]
   \right\}\,.
\end{eqnarray}
Since $S_B>S_F$, the first two terms in the curly brackets
dominate.
Thus, the tunneling amplitude $\Gamma$ of the false vacuum decay is 
found to be
\begin{equation}
   \Gamma\sim\frac{|\Psi(\xi_f)|^2}{|\Psi(0)|^2}
=\exp\left[-{1\over\hbar}(S_B-S_F)\right].
\end{equation}
This agrees with the result in the thin-wall limit.
The overall behavior of the wave function is shown
schematically in Fig.\ref{fig:wv2},
in which the normalization constant $N$ is chosen to
be $\exp[-(S_B-S_F)/2\hbar]$ so that the amplitude of
the wave function at the false vacuum equals unity.

\section{Discussion and conclusion}

We have presented a new method to calculate the tunneling
wave function that describes the false vacuum decay with gravity.
In this method, the tunneling wave function is constructed
by matching the false vacuum (FV) instanton and the Coleman-DeLuccia
(CD) instanton through the point at which the scale factor vanishes.
We found the resulting tunneling amplitude agrees with the naive
extrapolation of the formula whose validity had been
justified only in the thin-wall limit or in the flat background.
Our result is a strong support for the standard
interpretation that analytic continuation through
the critical bubble configuration of the
CD instanton describes the universe after false vacuum decay.

In our picture, the false vacuum decay with gravity becomes
more or less quantum cosmological. For example,
as can be seen from Fig.~\ref{fig:wv2}, interpreting
the wave function as describing the actual
tunneling path, the tunneling under
consideration is quite similar to a case of
standard quantum tunneling
with a resonant state inside the barrier, except for
the sign of the Euclidean actions $S_F$ and $S_B$
which are negative.
The difference is that the resonant state here is not
a state of something but the state of `nothing'.
This gives us a very interesting
picture that the false vacuum decay is first proceeded
by pumping up the amplitude in the state of nothing
and subsequently a universe with an expanding bubble is
created from nothing.

A similar idea of joining the two instantons was previously proposed
by Bousso and Chamblin in the context of path-integral approach\cite{BC}.
But the principles behind his approach and ours seem
rather different. Our method is completely within the
scope of the standard WKB approximation, while
they had to introduce a Planck size wormhole
and reverse the sign the 4-volume after traversing
the wormhole in an ad hoc way. Nevertheless, it is
of interest to see if these two methods correspond
to each other at certain level of approximations.

Recently, Rubakov and Sibiryakov proposed another method to deal with
the false vacuum decay with gravity\cite{RS}, by considering
complex paths and adding a constraint to the action to realize
the initial false vacuum state. It is, however,
not clear how the complex path they considered joins
the false vacuum and the critical bubble configuration,
particularly because of the constraint that modifies
the equations of motion.

\begin{figure}
\centerline{\epsfxsize=8cm\epsfbox{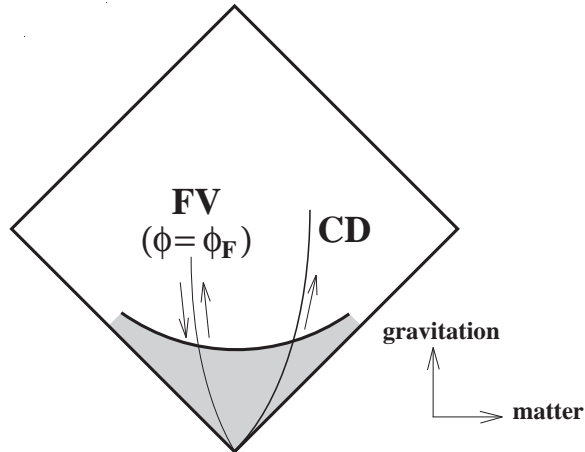}}
\caption{
Schematic picture of the causal structure of superspace
and the one-parameter family of configurations used
to construct the wave function.
The Euclidean region is shaded.
The lines indicated by FV and CD correspond to
the configurations of the false vacuum and Coleman-DeLuccia
solutions, respectively. The up or down arrows along the lines
indicate the expanding or contracting components contained
in the wave function.
}
\label{fig:ss}
\end{figure}

We have demanded the wave function to have
only the component describing an expanding universe after
false vacuum decay.
In this respect, our boundary condition is very
similar to the tunneling boundary condition proposed by
Vilenkin \cite{V} in the context of quantum cosmology.
However, in Vilenkin's picture, only the wave function
for an expanding universe appears in the Lorentzian
region of the superspace. This means there is
a steady flux coming out from the Euclidean boundaries
of the superspace. On the other hand, our wave function
is outgoing only for the universe after false vacuum decay.
Indeed, along the line $\phi=\phi_F$ in the superspace,
where the universe is in the false vacuum, our
wave function has both expanding and contracting components.
See Fig.~\ref{fig:ss}.
The latter feature is similar to the one of the
Hartle-Hawking wave function \cite{HH}.
These similarities of our wave function
with both the Vilenkin and Hartle-Hawking
wave functions can be seen also in Fig.~\ref{fig:wv2}:
If we consider only the right-hand side ($\xi>\pi/2$)
of the figure, it looks exactly like the cosmological
wave function satisfying the Vilenkin boundary condition.
In fact, the amplitude at $a=0$ ($\xi=\pi/2$)
is dominated by the component that decreases exponentially
as $a$ increases, as is the case of the Vilenkin wave function.
On the other hand, if we consider only the left-hand side
($\xi<\pi/2$) of the figure, our wave function can be
approximately regarded as a Hartle-Hawking wave function.
The exponentially decreasing
component becomes totally negligible when it appears
in the Lorentzian region, hence it is an irrelevant
component, while the exponentially increasing component
dominates the wave function in the Lorentzian region,
which consists of both outgoing and ingoing components
with equal weight. This is the characteristic
Hartle-Hawking feature.

We have considered only the leading order behavior of the WKB wave
function. We need to analyze the next order
to determine the quantum state of fluctuations
after tunneling. It seems, however, that there
is no reason to expect the result should differ from
the ones obtained previously
under the assumption of the Euclidean vacuum associated
with the CD instanton.
This is because all the properties of the quantum
fluctuations are determined by
the Euclidean structure of a single classical
solution, which is the CD instanton,
as far as we are interested only in the state
after false vacuum decay: The quantum fluctuations
around the CD instanton is completely decoupled from
those around the FV instanton.
Then once we focus only on the CD instanton,
our wave function satisfies the outgoing boundary condition,
and it is known to give
the Euclidean vacuum state\cite{VV}.
This reassures the validity of previous calculations of
quantum fluctuations in the one-bubble open inflation
scenario\cite{GMST}.

\bigskip
\centerline{\bf Acknowledgment}
We would like to thank J. Garriga and T. Tanaka
for useful discussions and comments at the
early stages of this work.
This work was supported in part by the Monbusho Grant-In-Aid
for Scientific Research, No.~09640355.
\appendix
\section{covariant formulation of multidimensional tunneling}
Here we recapitulate the covariant formulation of
the WKB wave function for multidimensional quantum tunneling
developed in \cite{TSY}, slightly adapted to the Wheeler-DeWitt
equation.

We consider the Hamiltonian in the form,
\begin{equation}
  \hat H = -{\hbar^2}g^{\alpha\beta}(q)
                     \frac1{f(q)}\V_\alpha f(q) \V_\beta +U(q)
\end{equation}
where $\{q^\alpha\}$ are the coordinates in the configuration
space of the dynamical variables, or superspace, and
$g_{\alpha\beta}$ is the superspace metric.
Expanding the wave function as
\begin{equation}
  \Psi=\exp\left[-\frac1\hbar(W_0+\hbar W_1+\cdots)\right],
\end{equation}
and inserting it into the Wheeler-DeWitt equation $\hat H\Psi=0$,
we obtain the lowest-order WKB equation,
\begin{equation}
  -\ g^{\alpha\beta}\V_\alpha W_0\V_\beta W_0 +U(q)=0,
\label{zeroWKB}
\end{equation}
and the first-order WKB equation,
\begin{equation}
  -g^{\alpha\beta}\V_\alpha W_0\V_\beta W_1
  +\frac12 g^{\alpha\beta}\V_\alpha\V_\alpha W_0
  +\frac12 g^{\alpha\beta}\frac{\V_\alpha f(q)}{f(q)}\V_\beta W_0=0.
\label{1stWKB}
\end{equation}
Introducing a parameter $\tau$ such that
\begin{equation}
  \frac{dq^a(\tau)}{d\tau}= g^{\alpha\beta}\V_\beta W_0\,,
\label{taudef}
\end{equation}
we obtain from Eq.~(\ref{zeroWKB}),
\begin{equation}
  W_0=\int d\tau\, U(q(\tau)),
\label{eq:defW0}
\end{equation}
and from Eq.~(\ref{1stWKB}),
\begin{equation}
  W_1 = \frac12 \ln
         \left[ 
           \det \left[\frac{\ppp q^\beta}{\ppp \lambda^\alpha}\right]
                \sqrt{|g|}f(q)
                     \right] +\mbox{const.},
 \label{eq:defW1}
\end{equation}
where $\{\lambda^{\alpha}\}=\{\tau,\lambda^n\}$ and $\{\lambda^n\}$
labels the different orbits of the congruence satisfying
 Eq.~(\ref{taudef}) on the superspace.

For our mini-superspace system (\ref{eq:WD}), 
the choice $\{q^\alpha\}=\{a,\phi\}$ leads to
\begin{equation}
   g_{\alpha\beta}=\left(\begin{array}{cc}
                          -a/\kappa & 0 \\
                           0 & 2\pi^2a^3
                         \end{array}\right)\,,
 \quad    f(q)=a^{p-1}. 
\label{eq:g-f}
\end{equation}
The congruence containing both FV and CD instantons
is expressed in the vicinity of $a=0$ as
\begin{equation}
   a(\tau,\phi_0)=\tau + O(\tau^3),
\quad
   \phi(\tau,\phi_0)=\phi_0 +O(\tau^2) .
\label{eq:a-phi}
\end{equation}
Hence a convenient choice of the label $\lambda$ of
the orbits is the value of $\phi$ at $\tau=0$, $\phi_0$.
Substituting Eqs.~(\ref{eq:g-f}) and (\ref{eq:a-phi}) 
into Eq.~(\ref{eq:defW1}), we find
\begin{equation}
  W_1 = \ln a^{\frac{p+1}{2}} + \mbox{const.}.  
\label{eq:W1}
\end{equation}
Equations (\ref{eq:defW0}) and (\ref{eq:defW1}) give
the WKB wave function to the first order,
\begin{eqnarray}
\Psi=N a^{-{p+1\over2}}\exp[-\int d\tau\,U(a(\tau),\phi(\tau))],
\end{eqnarray}
with $U=-(a/\kappa)Q(a,\phi)$.

\newcommand{\np}{Nucl. Phys. }

\end{document}